\documentclass[12pt,reqno]{article}
\usepackage{amsmath,hyperref,amssymb}
\usepackage{subfigure,graphicx,url}

\usepackage{tabularx}
\usepackage{bm}
\usepackage{euscript}
\usepackage{graphicx}
\usepackage{color}
\usepackage{amsfonts}
\usepackage{exscale}
\usepackage{amsbsy}
\usepackage{subfigure}
\usepackage{textcomp}

\newcommand{\be}{\begin{eqnarray}}
\newcommand{\ee}{\end{eqnarray}}

\newcommand{\bea}{\begin{eqnarray}}
\newcommand{\eea}{\end{eqnarray}}
\newcommand{\eeq}{\end{equation}}
\newcommand{\beq}{\begin{equation}}

\allowdisplaybreaks \numberwithin{equation}{section}

\DeclareSymbolFont{AMSa}{U}{msa}{m}{n}
\DeclareSymbolFont{AMSb}{U}{msb}{m}{n}
\DeclareMathSymbol{\fieldR}{\mathalpha}{AMSb}{"52}

\def\beq{\begin{equation}}
\def\eeq{\end{equation}}
\def\be{\begin{equation}}
\def\ee{\end{equation}}
\def\bea{\begin{eqnarray}}
\def\eea{\end{eqnarray}}

\newcommand{\ba}{\begin{eqnarray}}
\newcommand{\ea}{\end{eqnarray}}

\begin{document}
\begin{flushright} \small
\end{flushright}
\bigskip
\begin{center}
 {\large\bfseries Twining Genera of (0,4) Supersymmetric Sigma Models on K3 }\\[3mm]

 \

Sarah Harrison, Shamit Kachru and Natalie M. Paquette   \\[5mm]
 
 {\small\slshape
 SITP, Department of Physics\\
 and\\
 Theory Group, SLAC \\
 Stanford University\\
 Stanford, CA 94305, USA \\
\medskip
 {\upshape\ttfamily sarharr@stanford.edu, skachru@stanford.edu, npaquett@stanford.edu}
\\[3mm]}
\end{center}

\

\

\vspace{1mm}  \centerline{\bfseries Abstract}
\medskip
Conformal field theories with (0,4) worldsheet supersymmetry and K3 target can be used to 
compactify the $E_8 \times E_8$ heterotic string to six dimensions in a supersymmetric manner.
The data specifying such a model includes an appropriate configuration of 24 gauge instantons in the $E_8 \times E_8$ 
gauge group to satisfy the constraints of anomaly cancellation.
In this note, we compute twining genera -- elliptic genera with
appropriate insertions of discrete symmetry generators in the trace -- for (0,4) theories
with various instanton embeddings.   We do this by constructing linear sigma models which
flow to the desired conformal field theories, and using the techniques of localization.
We present several examples of such twining genera which are consistent with a moonshine relating these (0,4) models to 
the finite simple sporadic group $M_{24}$.

\bigskip
\newpage

\addtocontents{toc}{\protect\setcounter{tocdepth}{2}}

\tableofcontents



\section{Introduction}

To obtain a $(1,0)$ supersymmetric compactification of the heterotic string to six dimensions,
one should choose an internal CFT with (0,4) supersymmetry and right-moving central charge
$c_R = 6$.  In the realm of geometry, such CFTs arise as non-linear sigma models with K3 target.
In order to satisfy the Bianchi identify for the three-form field strength $H$ of the heterotic string
\begin{equation}
\label{Bianchi}
dH = {\rm Tr}(R\wedge R) - {\rm Tr}(F \wedge F)~,
\end{equation}
one should further embed 24 instantons into the $E_8 \times E_8$ gauge
group.  If one chooses bundles $V_{1,2}$ of rank $r_{1,2}$ in the two $E_8$s (which should be stable
and holomorphic, and have vanishing first Chern class $c_1 = 0$, in the simplest case), then the left-moving fermions in the sigma model couple to the gauge connections on these bundles, and $c_L = 4 + r_1 + r_2$.

The explicit construction of such (0,4) CFTs is a difficult task, and computations of observables in
such intricate theories are in general complicated to perform.  Ideally, one would like to be able to
compute the partition function of the internal conformal field theory.  But more generally, one has
to settle for obtaining coarser index information.  One such compromise is given by the elliptic
genus,

\begin{equation}
\label{elliptic}
Z(\tau,z) = {\rm Tr} (-1)^F y^{J_L} q^{L_0} \overline q^{\overline L_0},~~y = e^{2\pi i z}, ~q
= e^{2\pi i \tau}~.
\end{equation}

This is a graded trace over the Hilbert space of the left movers, containing further information 
about quantum numbers under a left-moving $U(1)$ current algebra whose generator is $J_L$.

In this paper, our focus will be on explicit examples of (0,4) models and their twining genera, which
are close relatives of the elliptic genus.
They can be defined as follows.  Consider a (0,4) theory with discrete symmetry $g$.  Then, one
can modify (\ref{elliptic}) to

\begin{equation}
\label{twining}
Z_g(\tau,z) = {\rm Tr} (-1)^F g~ y^{J_L} q^{L_0} \overline q^{\overline L_0}~,
\end{equation}

that is, one can take the trace with an insertion of the action of $g$ on the physical states.

We will construct (0,4) models as gauged linear sigma models with $K3$ target.  The basic
ideas involved in constructing such sigma models with Calabi-Yau target were developed in
the beautiful paper \cite{Witten}, and the extension to (0,2) models was discussed in detail in
\cite{Distler}.  As (0,4) models are a simpler subset of (0,2) models, our models will be simple
examples of the constructions in \cite{Distler}.  

We will compute the twining genera by using the techniques of localization. Localization was 
recently used to give a very explicit formula for the elliptic genus of linear sigma models with
rank one gauge groups in \cite{Benini}, with an extension to higher ranks appearing in 
\cite{Beninitwo}; a small modification of that formula suffices to compute
the twining genera (\ref{twining}).  Earlier results on the elliptic genera of (0,2) gauged 
linear sigma models
appeared in \cite{KawaiMohri}, which also anticipated (without derivation) aspects of the residue
formula of \cite{Benini}.

While one justification for computing the observables (\ref{twining}) is that they contain
valuable information about the spectrum of an interacting conformal field theory, we
had a more specific motivation for undertaking this study.  There is a Mathieu moonshine
relating the (4,4) sigma models with K3 target to the Mathieu group $M_{24}$ \cite{EOT}.  The
key first piece of evidence for this moonshine was a decomposition of the coefficients of the
q-expanded elliptic genus of K3,
in terms of dimensions of representations of $M_{24}$.  Given such a decomposition, one can make
predictions for the twining genera (\ref{twining}) for the (4,4) theories, if one inserts any element of $M_{24}$.  By
finding explicit realizations of symmetries of K3 sigma models, and computing (\ref{twining})
explicitly, one can check whether these symmetries correspond to (conjugacy classes of) elements
of the hypothetical $M_{24}$.  Such checks were carried out in \cite{MirandaTwining,
GaberdielTwining,EguchiTwining} with
impressive results.  The existence of a graded $M_{24}$ module with the desired properties has
since been proved at a rigorous level \cite{Gannon}.

Possible extensions of this moonshine to theories with only half as much supersymmetry, 
including (0,4) heterotic string compactifications, were discussed in \cite{Sixauthor}.  As a logical
extension of that work, it is desirable to find explicit symmetries of (0,4) K3 conformal field theories
and check if the twining genera (\ref{twining}) match with those of suitable $M_{24}$
conjugacy classes.  This note, as well as the companion \cite{toappear} to \cite{Sixauthor} which
studies twining genera of (0,4) supersymmetric K3 orbifold conformal field theories,
will present examples where exactly such matching can be seen.

\section{Some simple (0,4) gauged linear sigma models}

\subsection{Basic multiplets and terms in the action}

We will write down (0,4) linear sigma models by working in (0,2) superspace and using vector bundles constructed as the cohomology of an exact sequence, as in \cite{Distler}.  The enhanced worldsheet
supersymmetry is not manifest, but should be expected to emerge in the IR on general grounds when
we construct models which have a large-radius interpretation as K3 sigma models.  

The (0,2) multiplets we use are as follows (see \cite{Distler,Jacques} for more discussion).
(0,2) superspace has coordinates $(z,\bar z,\theta^+,\theta^-)$ (so $^\pm$ here on the
Grassman coordinates denotes $U(1)$ charge, not chirality).  The spinor superderivatives are
\begin{equation}
\overline D_{\pm} = {\partial \over \partial\theta^{\pm}} + \theta^{\mp}\partial_{\bar z}~.
\end{equation}
Chiral superfields $\Phi$ satisfy
\begin{equation}
\overline D_+ \Phi = 0
\end{equation}
and have a component expansion
\begin{equation}
\Phi = \phi + \theta^- \psi + \theta^- \theta^+ \partial_{\bar z} \phi
\end{equation}
with $\psi$ a right-moving fermion.  
Fermi superfields $\Lambda$ also satisfy $\bar D_+ \Lambda = 0$, but have component expansion
\begin{equation}
\Lambda = \lambda + \theta^- \ell + \theta^- \theta^+ \partial_{\bar z} \lambda
\end{equation}
instead, with $\lambda$ a left-moving fermion and $\ell$ an auxiliary field.

We will be considering (0,2) gauge theories with $U(1)$ gauge group, so we also need
to discuss the (0,2) gauge multiplet.  It consists of a pair of superfields $V, {\cal A}$ whose expansion,
in Wess-Zumino gauge, is given by
\begin{eqnarray}
&&V = \theta^- \theta^+ \bar{a} \nonumber \\
&&{\cal A} = a + \theta^+ \alpha - \theta^- \bar\alpha + {1\over 2} \theta^- \theta^+ D
\end{eqnarray}
with $a, \bar{a}$ the left/right moving pieces of the gauge field, $\alpha, \bar\alpha$ left-moving
gauginos, and D an auxiliary field.
The field strength supermultiplets are
\begin{eqnarray}
&&{\cal F} = -\alpha + \theta^- (D+f) - \theta^- \theta^+ \partial_{\bar z}\alpha \nonumber \\
&&\overline{\cal F} = - \bar\alpha + \theta^+ (D-f) + \theta^-\theta^+ \partial_{\bar z}\alpha
\end{eqnarray}
where 
\begin{equation}
f = 2 (\partial_z \bar a - \partial_{\bar z} a)~.
\end{equation}

The basic terms which appear in a supersymmetric action will be the following. 
A gauge invariant kinetic term for a charged chiral multiplet $\Phi$ with charge $Q$ is
\begin{eqnarray}
S_{\Phi} &=& \int d^2z (\partial_z - Qa) \bar\phi (\partial_{\bar z} + Q \bar a) \phi
+ (\partial_{\bar z} - Q \bar a)\bar\phi (\partial_z + Qa)\phi \nonumber  \\
&+&
2\bar\psi(\partial_z + Qa)\psi + Q(\bar\alpha \bar\psi\phi - \alpha \phi \bar\phi) - QD \bar\phi \phi~,
\end{eqnarray}
while a gauge invariant kinetic term for a charged Fermi multiplet $\Lambda$ of charge $Q$ is
\begin{equation}
S_{\lambda} = \int d^2z 2 \bar\lambda(\partial_{\bar z} + Q \bar a) \lambda - \bar \ell \ell~.
\end{equation}
The gauge kinetic
term is
\begin{eqnarray}
\label{gaugekinetic}
S_{\rm gauge} &=& -{1\over 2e^2} \int d^2z d^2\theta {\cal F} \overline{\cal F}\nonumber \\
&=& {1\over 2e^2} \int d^2z (f^2 - D^2 + 2\alpha \partial_{\bar z} \bar \alpha)~.
\end{eqnarray}
The Fayet-Iliopoulos D-term and $\theta$ angle are
\begin{equation}
S_{\rm D} = r \int d^2z D - i{\theta \over 2\pi} \int d^2z f
\end{equation}
(where $t = {\theta \over 2\pi} + ir$ plays the role of a K\"ahler parameter in large radius geometric
phases of the theories to come).
The (0,2) superpotential takes the form
\begin{eqnarray}
\label{superpot}
S_{\cal W} &=& \int d^2z ~d\theta^- \Lambda F(\Phi) + {\rm h.c.} \nonumber \\
&=& \int d^2z (\ell F(\phi) - \lambda {\partial F \over \partial \phi} \psi) + {\rm h.c.}~.
\end{eqnarray}
Here, $F$ needs to be chosen to be a homogeneous polynomial of the appropriate degree in
the charged field $\Phi$ so that (\ref{superpot}) is gauge invariant.

\subsection{The class of models of interest}

Our interest is to describe stable, holomorphic vector bundles $V$ with $c_1(V) = 0$ and
$c_2(V) = c_2(TM)$ over K3 surfaces $M$.  A simple class of models which admits a gauged
linear sigma model description is the following.  We choose for $M$ the 
Calabi-Yau hypersurface in the
$W{\mathbb P}^{3}$ with weights $w_i$ ($i=1,...,4$), described by the equation
\begin{equation}
\label{hyper}
W(\phi_i) = 0 \subset W{\mathbb P}^{3}_{w_1,...,w_4}~.
\end{equation}
We define $V$ as the cohomology of the exact sequence
\begin{equation}
\label{bundle}
0 \to V \to \oplus_a {\cal O}(n_a) \stackrel{\otimes F_a(\phi)}{\to} {\cal O}(m) \to 0~.
\end{equation}
The conditions that $c_1(TM) = c_1(V)=0$ and $c_2(TM) = c_2(V)$ are captured by the
Diophantine equations
\begin{eqnarray}
\label{dioph}
&&\sum_i w_i = d,~~\sum_a n_a = m,\nonumber\\
&& m^2 - \sum_a n_a^2 = d^2 - \sum_i w_i^2~,
\end{eqnarray}
with $d$ being the degree of the defining polynomial $W(\phi)$ of the K3 surface;
these equations follow simply from the adjunction formula for Chern classes.
The second equation in (\ref{dioph}) just imposes the requirement of worldsheet gauge anomaly cancellation
for the abelian gauge field.

These theories can be represented as gauged linear sigma models in the following way.
Let us consider the (0,2) supersymmetric abelian gauge theory with the matter content shown in Table 1.

\begin{table}[ht]
\caption{Field content of our (0,2) sigma models}
\centering
\begin{tabular}{c c}
\\
Field & Gauge charge \\ 
\hline
$\Phi^i$ & $w_i$ \\
$P$ & $-m$\\
$\Lambda^a$ & $n_a$ \\
$\Gamma$ & $-d$ \\ 
\hline
\end{tabular}
\label{table:charges}
\end{table}

$\Phi^i, P$ are (0,2) chiral multiplets, while $\Lambda^a$ and $\Gamma$ are Fermi multiplets.
For our (0,2) superpotential we choose
\begin{equation}
\int d^2z ~d\theta^- (\Gamma W(\Phi) + P \Lambda^a F_a(\Phi)) + {\rm h.c.}
\end{equation}
with $W$, $F_a$ coinciding with the data in the definition of the K3 hypersurface and the bundle $V$ above.
One can verify, as in \cite{Distler}, that in the limit of large $r$, this theory flows to the sigma model governed by
the geometric objects (\ref{hyper}) and (\ref{bundle}), with the scalars living on the hypersurface 
(\ref{hyper}) while the left-moving fermions transform as sections of the bundle (\ref{bundle}).
Of course, as one varies the Fayet-Iliopoulos parameters in such a gauge theory, other interesting phases can
arise (with Landau-Ginzburg orbifold phases being a prototypical such phase).

We will need to generalize this construction in a trivial way, in order to capture the geometry of two
non-trivial bundles $V_{1,2}$ which we embed
into the two $E_8$s.  The appropriate generalization is to introduce two chiral analogues of the $P$ field $P^{1,2}$, with charges $m_{1,2}$, and two sets of
Fermi multipets $\Lambda_1^a$ and $\Lambda_2^\alpha$ of charges $n_a$ and $q_{\alpha}$, with $a = 1,...,r_1 + 1$ and $\alpha = 1,...,r_2+1$.
The superpotential is now
\begin{equation}
\int d^2z ~d\theta^- (\Gamma W(\phi) + P_1 \Lambda_1^a F_a(\Phi) + P_2 \Lambda_2^\alpha G_{\alpha}(\Phi))~,
\end{equation}
with $F_a$ and $G_{\alpha}$ defining the bundles $V_{1,2}$ through exact sequences as in (\ref{bundle}).
The constraints on the Chern classes now become
\begin{eqnarray}
\label{cclasses}
&&m_1 = \sum n_{a}, m_2 = \sum q_{\alpha} \nonumber \\
&&d^2 - \sum w_i^2 = (m_1^2 - \sum n_a^2) + (m_2^2 - \sum q_{\alpha}^2)~.
\end{eqnarray}
Again, the second equation in (\ref{cclasses}) is required for gauge anomaly cancellation, and is interpreted in
space-time as implementing the condition 
\begin{equation}
\label{ctwocond}
c_2(TM) = c_2(V_1) + c_2(V_2)~, 
\end{equation}
which is required to satisfy the 
Bianchi identity (\ref{Bianchi}).

Intuitively, the equation (\ref{ctwocond}) means that in perturbative supersymmetric heterotic models on K3, one should choose non-negative integers
$n^{(1)}, n^{(2)}$ with 
\begin{equation}
n^{(1)} + n^{(2)} = 24~,
\end{equation}
and place $n^{(1)}$ and $n^{(2)}$ gauge instantons in the two $E_8$s.  Our goal in the next
section will be to show that
in a variety of examples constructed as above, reflecting distinct choices of $n_{1,2}$, one can find (0,4)
sigma models with discrete symmetries $g$ whose twining genera (\ref{twining}) are consistent with the properties expected
from Mathieu moonshine for (0,4) models.  This strengthens the case made in \cite{Sixauthor} that moonshine extends to
a portion of the web of 4d ${\cal N}=2$ (or 6d ${\cal N}=1$) supersymmetric heterotic string theories, as well as their
type II (or F-theory) Calabi-Yau duals.

\subsection{Specific examples of models and discrete symmetries}

We will focus on four classes of specific models with different values of $n^{(1)}$ and $n^{(2)}$, but it should be clear that many other models 
exist and could be fruitfully analyzed in this way.   In each case, we just discuss some simple symmetries which arise
for easy choices of the defining data; we are not exhaustive.  We label the models by the
instanton numbers $(n^{(1)},n^{(2)})$ chosen in each.  The four models we will study are:

\subsubsection{Model 1: A (24,0) model}

For our first example, we will study the theory with $d=4$ and ${w_i}={1,1,1,1}$.   To begin with, we can choose the defining data of the target manifold to be
\begin{equation}
\label{Fermat}
W(\Phi_i) = \sum_i {1\over 4} \Phi_i^4~,
\end{equation}
i.e. the Fermat point in the moduli space of this K3 hypersurface.  The bundle is defined by choosing
\begin{equation}
V_1~:~m=4, ~\{n_a\} = \{1,1,1,1\},~F_{a}(\phi) = \Phi_{a}^3~.
\end{equation}
For generic defining data, this model simply defines the $(0,4)$ model obtained by deforming the tangent bundle
of $K3$ away from the (4,4) supersymmetric locus, while extending the rank from $SU(2)$ to $SU(3)$ (by partially
Higgsing the $E_7$ space-time gauge group with a ${\bf 56}$ of $E_7$).  
It has instanton numbers $(n^{(1)}=24, n^{(2)} = 0)$.
The (4,4) theory was studied in more detail
in \cite{GabSym}; we simply discuss this model here to provide a warm-up on more or less familiar territory.

We can study several simple symmetries in the Fermat K3.  We will study three:\\
\medskip
\noindent
1. The $\mathbb Z_2$ symmetry which acts as
\begin{equation}
g~:~\Phi_{1,2} \to -\Phi_{1,2},~~\Lambda_{1,2} \to - \Lambda_{1,2}~.
\end{equation}
\medskip
\noindent
2.  The $\mathbb Z_3$ symmetry which acts as a permutation of cycle shape $(123)$ on $\Lambda_{1,2,3}$ and $\Phi_{1,2,3}.$
\medskip
\noindent
3.  The $\mathbb Z_4$ symmetry 
\begin{equation}
g~:~\Phi_{1,2} \to \pm i \Phi_{1,2},~\Lambda_{1,2} \to \pm i \Lambda_{1,2}~.
\end{equation}

We can also obtain more elaborate symmetries by choosing slightly different data.  For instance, if we choose a complex structure
\begin{equation}
W(\Phi_i) = \Phi_1^3 \Phi_2 + \Phi_2^3 \Phi_3 + \Phi_3^3 \Phi_4 + \Phi_4^3 \Phi_1
\end{equation}
then we can find a $\mathbb Z_5$ symmetry:\\
\medskip
\noindent
4. $\mathbb Z_5$ symmetry:
\begin{eqnarray}
g~:~&&\Phi_1 \to \lambda \Phi_1,~\Phi_2 \to \lambda^2 \Phi_2, ~\Phi_3 \to \lambda^4 \Phi_3, ~\Phi_4 \to \lambda^3 \Phi_4,
~~\lambda \equiv e^{2\pi i \over 5}~,\nonumber\\
&&\Lambda_1 \to \lambda \Lambda_1, ~\Lambda_2 \to \lambda^2 \Lambda_2,~\Lambda_3 \to \lambda^4 \Lambda_3,
~\Lambda_4 \to \lambda^3 \Lambda_4~.
\end{eqnarray}
Defining data for the vector bundle which respects this symmetry could include e.g. $F_a(\Phi) = {\partial W \over \partial \Phi_a}$
or suitable variants.

Another K3 which admits an interesting symmetry has the complex structure
\begin{equation}
W(\Phi_i) = \Phi_1^3 \Phi_2 + \Phi_2^3 \Phi_3 + \Phi_3^3 \Phi_1 + \Phi_4^4~.
\end{equation}
This surface admits the $\mathbb Z_7$ symmetry:

\medskip
\noindent
5. $\mathbb Z_7$ symmetry:
\begin{eqnarray}
g~:~&&\Phi_1 \to \lambda \Phi_1,~\Phi_2 \to \lambda^4 \Phi_2,~\Phi_3 \to \lambda^2 \Phi_3,~~\lambda \equiv e^{2\pi i \over 7}~,
\nonumber \\
&&\Lambda_1 \to \lambda \Lambda_1, ~\Lambda_2 \to \lambda^4 \Lambda_2,~\Lambda_3 \to \lambda^2 \Lambda_3.
\end{eqnarray}
Again suitable defining data for the bundle could be $F_a(\Phi) = {\partial W \over \partial \Phi_a}$ with other choices also possible.

\subsubsection{Model 2: A (12,12) model }

Again reverting to the Fermat quartic K3 (\ref{Fermat}), we choose now bundles $V_{1,2}$ each with
$m_{1,2}=3$ and $ \{n_a\}, \{q_\alpha\} = \{ 1,1,1 \}$.   We consider the symmetries:\\
\medskip
\noindent
1.  A $\mathbb Z_2$ with
\begin{equation}
g~:~\Lambda_{1,2,3} \to -\Lambda_{1,2,3},~P_1 \to -P_1,~\Phi_{1,2} \to -\Phi_{1,2}~.
\end{equation}
Here the $\Lambda$s are those spanning $V_1$, and one should choose data $F_a(\phi)$ which is consistent with
the symmetry.\\
\medskip
\noindent
2. A $\mathbb Z_4$ with
\begin{equation}
g~:~ \Lambda_{1,2} \to \pm i \Lambda_{1,2},~\Phi_{1,2} \to \pm i \Phi_{1,2}~.
\end{equation}
Again, these fermions are from $V_1$, and one should choose data $F_{1,2}(\phi)$ consistent with the symmetry.

\subsubsection{Model 3: A (14,10) model}

Now, we work on the K3 hypersurface embedded in $W{\mathbb P}^3_{1,1,1,3}$.  For a defining equation, we choose
\begin{equation}
\label{sextic}
W(\Phi) = \Phi_{1}^6 + \Phi_2^6 + \Phi_3^6 + \Phi_4^2~
\end{equation}
For bundles, we let $V_1$ be specified by $m_1 = 5, \{ n_a \} = \{ 3,1,1 \}$ and
$V_2$ be specified by $m_2 = 4, \{ q_\alpha \} = \{ 2,1,1 \}.$  

We consider two symmetries in this model:\\
\medskip
\noindent
1. A representative $\mathbb Z_2$ symmetry is, for instance,
\begin{equation}
g~:~\Lambda_{2,3} \to - \Lambda_{2,3}, ~\Phi_{2,3} \to -\Phi_{2,3}~,
\end{equation}
with the $\Lambda$s being fermions involved in the construction of $V_1$.  Simple choices of
the $F_a(\Phi)$ are consistent with such a symmetry.

\medskip
\noindent
2.  We can consider a $\mathbb Z_3$ symmetry as follows:
\begin{equation}
g~:~\Phi_{1} \to e^{2\pi i \over 3} \Phi_1, ~\Phi_2 \to e^{4\pi i \over 3}\Phi_2
\end{equation}
with the two charge $1$ fermions in $V_1$, $\Lambda_{2,3}$, rotating as
\begin{equation}
g~:~\Lambda_2 \to e^{4\pi i \over 3}\Lambda_2,~\Lambda_3 \to e^{2\pi i \over 3}\Lambda_3~.
\end{equation}
There are simple choices of the $F_a(\Phi)$ that accomodate this symmetry.

\subsubsection{Model 4: An (18,6) model}

Finally, still working on the K3 hypersurface (\ref{sextic}), we study the bundles
$V_1$ with $m_1=5, \{ n_a \} = \{ 2,1,1,1\}$ and $V_2$ with $m_2 = 3$, $\{ q_\alpha \} = \{ 1,1,1 \}.$
One $\mathbb Z_3$ symmetry arises in this model by permuting the fermions $\Lambda_{2,3,4}$ of charge 1
arising as part of $V_1$; the fermions $\Lambda_{4,5,6}$ arising as part of $V_2$; and the chiral fields
$\Phi_{1,2,3}$, all with the permutation of cycle shape $(123)$.    Once again, simple choices of the bundle data $F(\Phi)$ are consistent with such a symmetry.

\section{Computation of the twining genera}

In this section, we compute the twining genera under the various model symmetries described in \S2.3.  We begin by
outlining the general strategy and formulae that are relevant, and then simply present the results of applying these
formulae to the various cases.  Our work relies heavily on the elegant residue formula derived recently in
\cite{Benini}.

\subsection{Residue formula for elliptic genus}

The elliptic genus was first discussed in \cite{SW,Pilch,Edold}.  Its application to string compactification was
pioneered in \cite{EOTY}, and it was first computed by localization in (2,2) supersymmetric Landau-Ginzburg models in
\cite{EdLG} and for (0,2) models in \cite{KawaiMohri}.  It has recently been the focus of attention in, for instance, \cite{Benini,
Gukov,Beninitwo}

The formalism we discuss only assumes $\mathcal N=(0,2)$ supersymmetry, though our application will be to
$(0,4)$ theories .  Although in many discussions of the elliptic genus in theories with $(2,2)$ supersymmetry the left-moving
R-symmetry plays a crucial role, here there is no longer a left-moving R-charge.  However, the models we consider will
have an extra $U(1)$ global current $J_L$, and we will grade by the quantum number under the associated charge in the elliptic genus. In the models described in the previous section, $J_L=0$ for $\Gamma$ and  $\Phi_i$ , and for the $\Lambda_{a,\alpha}$, $J_L=-1$, whereas for the $P_{a,\alpha}$, $J_L=+1$. 

We follow the discussion of $(0,2)$ abelian gauge theory in \cite{Benini}.  Let us define $u$ to be the holonomy of the $U(1)$
gauge field around the cycles of the torus
\begin{equation}
u = \oint A_t dt ~-~\tau \oint A_s ds
\end{equation}
with $t,s$ the temporal and spatial directions, and $\tau$ the modular parameters of the torus.
The elliptic genus is given by the graded trace
\begin{equation}
\label{whatwewant}
Z(\tau,z) =  {\rm Tr}_{\rm RR} (-1)^F y^{J_L} q^{H_L} \bar q^{H_R} ~.
\end{equation}
Obtaining a formula for (\ref{whatwewant}) via localization involves doing an integral over the Wilson lines
$u$ of the abelian gauge field.

This integral localizes to a sum of contour integrals around loci (in the moduli space of flat connections) where some of the
fields become massless; we refer to these as singular points. Let us consider a general $(0,2)$ $U(1)$ gauge theory, with a number of gauge charged chiral and Fermi multiplets $\Phi_i$ and $\Lambda_a$, as well as one vector multiplet.
Suppose that the charges of the chiral and Fermi multiplets under the gauge and $U(1)$ global symmetry are $Q_{i,a}$ and $J_{i,a}$ respectively.
Then, defining
\be
x = e^{2\pi i u}~,
\ee
the expression that has been obtained for the elliptic genus is \cite{Benini}
\be
\label{contourformula}
Z(\tau,z)=- {\eta(q)^2}\sum_{u_j\in \mathcal M^+}\oint\limits_{u=u_j}du \prod_{\Phi_i}\frac{i\eta(q)}{\theta_1(q,y^{J_i}x^{Q_i})}\prod_{\Lambda_a}\frac{i\theta_1(q,y^{J_a}x^{Q_a})}{\eta(q)},
\ee
where $\mathcal M^+$ is the relevant set of singular points.\footnote{Our conventions for modular forms can be found in appendix A.} These points are defined as the solutions to the equation
\be
Q_iu+J_iz\equiv 0 \mod (\mathbb Z+\tau\mathbb Z),
\ee 
with ${\it positive}$ $Q_i$.
Equivalently one could sum over poles in the set $\mathcal M^-$ (including an overall change of sign, due to the reversed orientation of the contour),  defined by solutions to the above equation for all 
${\it negative}$ $Q_i$.

One can roughly understand the origin of the formula (\ref{contourformula}) as follows.  Each chiral, Fermi and vector multiplet
makes a (multiplicative) contribution to the index at any fixed value of the Wilson lines $u$. 
For a (0,2) chiral multiplet with global $U(1)$ charge $J$ and flavor charge $Q$, the contribution is
\be
Z_{\Phi,J,Q}^{(0,2)}(\tau,z,u)=\frac{i\eta(q)}{\theta_1(q,y^{J}x^Q)}~.
\ee
That of a Fermi multiplet with global $U(1)$ charge $J$ is
\be
Z_{\Lambda,J,Q}(\tau,z,u)=\frac{i\theta_1(q,y^{J}x^Q)}{\eta(q)}.
\ee
Finally, the contribution of a (0,2) vector multiplet is
\be
Z_{\rm vector}^{(0,2)}(\tau)=\eta(q)^2.
\ee
independent of $u$.  The product of these expressions over all multiplets present in a given theory, integrated over the $u$-plane,
can be reduced to the formula (\ref{contourformula}).

\subsubsection{K3 elliptic genus}

The standard results for the elliptic genus of K3 (or in the language of quantum field theory,
for the $\mathcal N=(4,4)$ sigma model with K3 target) is \cite{EOTY}
\be
Z_{K3}(\tau,z)=8\sum_{i=2}^4\left ({\theta_i(q,y)\over \theta_i(q,1)}\right )^2,
\ee
which has the expansion
\be
Z_{K3}\sim \left ({2\over y}+20+2y\right )+\left ({20\over y^2}-{128\over y}+216-128y+20y^2\right)q+\ldots
\ee

For a $(0,4)$ model on K3 with rank $r$ gauge bundle, the elliptic genus is given by 
\be
\label{universal}
Z_{K3}^{r}=\left( \frac{\theta_1(q,y)}{i\eta(q)}\right )^{r-2}Z_{K3},
\ee
as derived in \cite{KawaiMohri}.  It is easy to check that applying (\ref{contourformula}) to our models
of \S2.3\ agrees with the result (\ref{universal}), with $r = r_1 + r_2$ the sum of the ranks of the bundles
embedded in the two $E_8$s.

\subsection{Residue formula for twining genera}

Our real interest is to compute the elliptic genus with the insertion of a symmetry operator, $g$, into the path integral
\be
Z^{(n_1,n_2)}_{g}(\tau,z)={\rm Tr}_{\rm RR} ~g~(-1)^F y^{J_L} q^{H_L}\bar q^{H_R}
\ee
for various particular $(n_1,n_2)$ instanton embeddings. We can do this with a slight modification to the computation of the untwined elliptic genus. 

Consider an operator $g$ which acts on chiral and Fermi multiplets as 
\be
g\Phi_i=e^{2\pi i\alpha_i}\Phi_i,~~g\Lambda_a=e^{2\pi i\beta_a}\Lambda_a,
\ee
and is a symmetry of the action. When inserting this operator into the path integral, it modifies the contribution due to the chiral and Fermi multiplets. The contribution of a $(0,2)$ chiral multiplet $\Phi_i$ to the integrand in  (\ref{contourformula}) becomes 
\be
\frac{i\eta(q) ~e^{\pi i\alpha_i}}{\theta_1(q,e^{2\pi i\alpha_i}y^{J_i}x^{Q_i)}}~,
\ee
while one obtains 
\be
i \theta_1(q, e^{2\pi i\beta_a}y^{J_a} x^{Q_a}) \over e^{i\pi\beta_a} ~\eta(q)
\ee
from the twined Fermi multiplet $\Lambda_a$.  One then sums over the (now shifted) poles that
previously contributed to the elliptic genus - the detailed locations of the poles in $\mathcal M^+$ on
the $u$-plane, as well
as their orders, can be modified depending on the $g$ charges of the fields involved.

Denote the elliptic genus of the $(4,4)$ theory twined by a conjugacy class $g$ of $M_{24}$ by $Z_g$.\footnote{
The $Z_g$ are discussed in detail in appendix B, where also the $M_{24}$ character table and the
first few coefficients in the q-expansion of the various $Z_g$ are presented.}
Then we expect the twined elliptic genus of an $(n_1,n_2)$ model to decompose as
\be
\label{assume}
Z_g^{(n_1,n_2)}=\text{ch}(SO(2r-4))Z_{g},
\ee
i.e. a product of twined $(4,4)$ genera and twined $SO(2r-4)$ characters. 

In writing (\ref{assume}), we are making two important assumptions:

\noindent
1) We assume that the $M_{24}$ module which is relevant in the moonshine for $(0,4)$ models with arbitrary instanton embeddings, has the same representations at each level as the one which arises in the $(4,4)$ theory.  Evidence for this was presented already
in the new supersymmetric index computations of \cite{Sixauthor}, which are valid for all instanton embeddings.

\noindent
2) We are assuming that the factor of
\be
\left( \frac{\theta_1(q,y)}{i\eta(q)}\right )^{r-2}
\ee
in the elliptic genus of a $(0,4)$ theory with rank $r$ bundle transforms as an element of the spinor minus conjugate spinor representation of $SO(2r-4)$.  This is motivated by the results to appear in the companion paper about $(0,4)$ orbifolds
\cite{toappear}.  Heuristically, the $SO(2r-4)$ symmetry could appear manifestly in a field theory where one deformed the
bundle $V_1 \oplus V_2$ to be an $SU(2)$ bundle with instanton number $n^{(1)} + n^{(2)}$.  As the elliptic genus is 
invariant under such smooth deformations, this may explain the appearance of such factors (related to further `hidden
symmetries') in the twining genera of $(0,4)$ sigma models.

We now show that our results for the set of models discussed in \S2.3\
satisfy the assumption (\ref{assume}).  We view this as a check of $M_{24}$ moonshine for $(0,4)$ theories with a variety of
instanton embeddings.

\subsection{Examples}

\subsubsection{Model 1}

Here, we considered five symmetries in \S2.3.1: a $\mathbb Z_2$ symmetry, a $\mathbb Z_3$ symmetry, a $\mathbb Z_4$ symmetry, a $\mathbb Z_5$ symmetry
and a $\mathbb Z_7$ symmetry.
The results for the twining genera are:
\begin{eqnarray}
&&Z_{\mathbb Z2} = {\theta_1(y) \over i\eta(q) } Z_{2A}~,\nonumber\\
&&Z_{\mathbb Z3} = {\theta_1(y) \over i\eta(q) } Z_{3A}~,\nonumber\\
&&Z_{\mathbb Z4} = {\theta_1(y) \over i\eta(q) } Z_{4B}~,\\
&&Z_{\mathbb Z5} = {\theta_1(y) \over i\eta(q)} Z_{5A}~,\nonumber\\
&&Z_{\mathbb Z7} = {\theta_1(y) \over i\eta(q)} Z_{7A}~.\nonumber
\end{eqnarray}
Here, $Z_{2A}, Z_{3A}$, $Z_{4A}$, $Z_{5A}$ and $Z_{7A}$ are the corresponding twining genera of the $(4,4)$ elliptic
genus with an insertion in those $M_{24}$ conjugacy classes (see appendix B). The first argument of the theta function has been suppressed here and below.

\subsubsection{Model 2}

We considered two symmetries in \S2.3.2: a $\mathbb Z_2$ symmetry and a $\mathbb Z_4$ symmetry.  The results for the twining genera are:
\begin{eqnarray}
&&Z_{\mathbb Z2} = {\theta_1(y)^2 \over (i\eta(q) )^2} Z_{2A}~,\nonumber\\
&&Z_{\mathbb Z4} = {\theta_1(iy) \theta_1(-iy) \over (i\eta(q) )^2} Z_{4B}~.
\end{eqnarray}

\subsubsection{Model 3}

Here, we also considered two symmetries in \S2.3.3 -- a $\mathbb Z_2$ and a $\mathbb Z_3$.  The results are:
\begin{eqnarray}
&&Z_{\mathbb Z2} = \left(\theta_1(y) \over i\eta(q) \right)^2 Z_{2A}~,\nonumber\\
&&Z_{\mathbb Z3} = {\theta_1(e^{2\pi i \over 3}y) \theta_1(e^{4\pi i \over 3}y)\over (i\eta(q))^2}
~Z_{3B}~.
\end{eqnarray}

\subsubsection{Model 4}

We considered a $\mathbb Z_3$ symmetry in \S2.3.4.  The result is
\begin{equation}
Z_{\mathbb Z3} = {{\theta_1(y) \theta_1(e^{2\pi i\over 3}y) \theta_1(e^{4\pi i \over 3}y)} \over
(i\eta(q))^3} Z_{3A}~.
\end{equation}

\section{Discussion}

In this note, we used the recently derived localization formula for the elliptic genus of $(0,2)$ supersymmetric rank one two-dimensional gauge
theories \cite{Benini} to compute twining genera of (0,4) gauged linear sigma models with K3 target.  We did this for a variety
of discrete symmetries in $(0,4)$ models with four different sets of instanton numbers $(n^{(1)},n^{(2)})$.

In several cases, we found that the simple discrete symmetries give twining genera which are consistent with those of 
$M_{24}$ elements of the same order, with the trace in the elliptic genus taken over the $M_{24}$ module conjectured
to exist in \cite{EOT} and constructed in \cite{Gannon}.  These direct computations are an analogue, for a conjectural (0,4) moonshine with various instanton
numbers, of the twining calculations in \cite{MirandaTwining,GaberdielTwining,EguchiTwining}.  
Interestingly, the 3B conjugacy class of $M_{24}$, which does not descend from the classical symmetries of K3 surfaces
(as they lie in $M_{23}$ \cite{Mtwentythree}) and which has been elusive, appears here in one of the first cases we
examined.  

It should not be difficult to find linear sigma models which admit relatively elaborate discrete symmetries.  The $\mathbb Z_5$
and $\mathbb Z_7$ examples of \S2.3.1\ were found by using a strategy developed in \cite{Greeneold}, and it seems quite plausible
that one can write down examples which show twining in higher order $M_{24}$ conjugacy classes in this way.  It should also be instructional to go through the list of e.g. the `famous 95' weighted projective K3 hypersurfaces of Reid \cite{Reid}, and see which of them admit interesting symmetries;
this may lead to interesting new examples even in the $(4,4)$ theory.

A major question which remains is the proper interpretation of the evidence presented here, as well as in
\cite{Sixauthor,toappear}, for a moonshine relating heterotic $(0,4)$ theories (and their type II Calabi-Yau duals) to $M_{24}$.
The observations of \cite{GabSym} indicate that $M_{24}$ does not play a canonical role as an embedding group for
symmetries of $(4,4)$ superconformal theories with K3 target.  The symmetries available in $(0,4)$ theories will of course
only be richer; developing a classification would be very interesting.  Failing a complete classification, a detailed study of particular families with large symmetry groups (extending the philosophy of 
\cite{Wendland} from the $(4,4)$ case) could also prove illuminating.  It is even within the realm of possibility that some
$(0,4)$ superconformal theory, or perhaps a non-perturbative heterotic vacuum with small instantons replacing the gauge
bundles $V_{1,2}$, could manifest the full symmetry and `explain' the appearance of $M_{24}$ in the elliptic genus.
But other interpretations of the moonshine, in terms of Rademacher sums arising naturally in AdS/CFT \cite{JohnMiranda},
or in terms of supersymmetric indices of NS5 branes \cite{Jeff}, are also quite promising.  Related
directions to explore are discussed in \cite{Govindarajan,Yang}.

\bigskip
\centerline{\bf{Acknowledgements}}

We are grateful to Francesco Benini for helpful discussions about the results in \cite{Benini}, and
to Tohru Eguchi and Kazuhiro Hikami for permission to use their tables of $M_{24}$ characters
and twining genera in appendix B.
We are also very happy to thank Miranda Cheng, Xi Dong, John Duncan, Jeff Harvey, Daniel Whalen and Timm Wrase for extensive
discussions about twining genera and moonshine in general.
S.H. and S.K. enjoyed the hospitality of the Simons Center workshop on ``Mock Modular Forms,
Moonshine and String Theory" as this work was completed, and thank the participants for creating
a very stimulating intellectual atmosphere.
  S.K. acknowledges the support of the NSF
under grant PHY-0756174, the Department of Energy under contract DE-AC02-76SF00515, and the John Templeton
Foundation.  S.H. is supported by the John Templeton Foundation, and N.P. by a Stanford Humanities and Sciences
Fellowship.

\appendix

\section{Conventions}\label{sec:conventions}
We use the following conventions for the Jacobi $\theta_i(q,y)$ functions
\ba
\theta_1(q,y) &=& i \sum_{n=-\infty}^{\infty} (-1)^n q^{\frac{(n-\frac12)^2}{2}} y^{n-\frac12}\,,\\
\theta_2(q,y) &=& \sum_{n=-\infty}^{\infty} q^{\frac{(n-\frac12)^2}{2}} y^{n-\frac12}\,,\\
\theta_3(q,y) &=& \sum_{n=-\infty}^{\infty} q^{\frac{n^2}{2}} y^n\,,\\
\theta_4(q,y) &=& \sum_{n=-\infty}^{\infty} (-1)^n q^{\frac{n^2}{2}} y^n\,,
\ea
where $q=e^{2 \pi i \tau}$ and $y=e^{2 \pi i z}$. Whenever the $y$-dependence is not specified, we have set $y=1$, for example $\theta_i = \theta_i(q) = \theta_i(q,1)$ and likewise for the other functions defined below.
These $\theta_i(q,y)$ functions have the following product expansion
\ba
\theta_1(q,y) &=& -i q^{1/8} y^{1/2} \prod_{n=1}^{\infty} (1-q^n)(1-y q^n)(1-y^{-1}q^{n-1})\,,\\
\theta_2(q,y) &=& q^{1/8} y^{1/2} \prod_{n=1}^{\infty} (1-q^n)(1+y q^n)(1+y^{-1}q^{n-1})\,,\\
\theta_3(q,y) &=& \prod_{n=1}^{\infty} (1-q^n)(1+y q^{n-\frac12})(1+y^{-1}q^{n-\frac12})\,,\\
\theta_4(q,y) &=& \prod_{n=1}^{\infty} (1-q^n)(1-y q^{n-\frac12})(1-y^{-1}q^{n-\frac12})\,.
\ea

We also use the Dedekind $\eta(q)$ function
\be
\eta(q) = q^\frac{1}{24} \prod_{n=1}^\infty (1-q^n).
\ee

\section{$M_{24}$ character table and coefficients of twining genera}

In \S3, we expressed the results for twining genera in various (0,4) models in terms of the 
$Z_{g}$ which appear in the twined elliptic genus of the (4,4) K3 sigma model, for various $M_{24}$
conjugacy classes $g$.  In practice, to work out the q-expansions for the resulting forms, one
needs the character table of $M_{24}$.  It is reproduced in Table 2 for completeness.
The classes appearing before 12B in the top row can also be considered as conjugacy classes in 
$M_{23}$, while 12B and those appearing to its right are intrinsic elements of $M_{24}$ with no
precursor in $M_{23}$.

\begin{table}[h!]
\begin{center}
\includegraphics[width=0.9\textwidth]{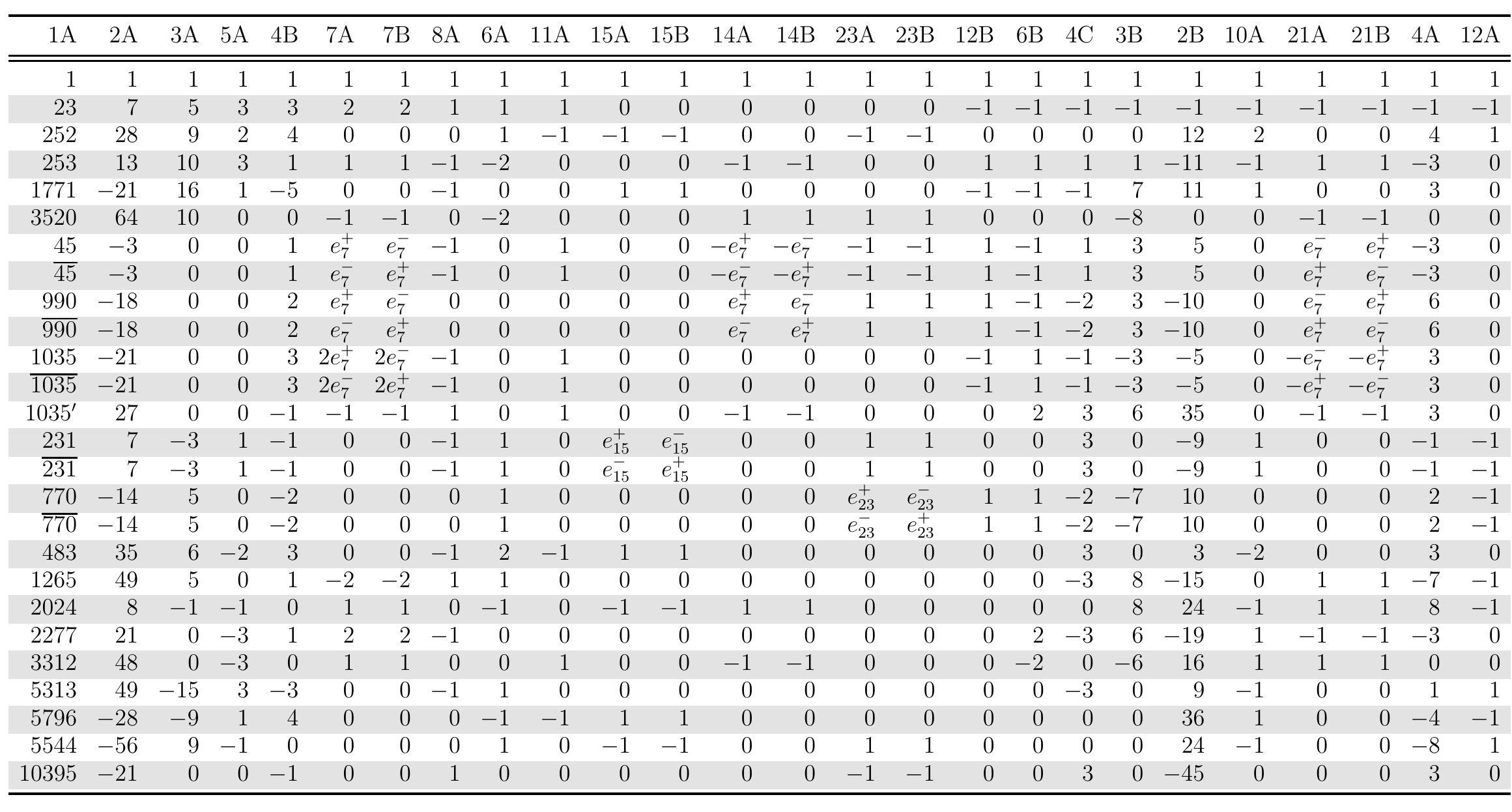}
\end{center}
\caption{Character table for $M_{24}$.}
\end{table}

The q-expansions of the $Z_g$ can be written as follows.  For the elliptic genus of K3, one writes
\begin{equation}
Z_{K3}(z;\tau) = 20 {\rm ch}_{h=1/4, \ell = 0}(z;\tau) -
2{\rm ch}_{h=1/4, \ell=1/2}(z;\tau) + \sum_{n=1}^\infty A(n) {\rm ch}_{h=n+1/4,\ell=1/2}(z;\tau)
\end{equation}
where ${\rm ch}_{h,\ell}$ are characters of the ${\cal N}=4$ superconformal algebra with a given
conformal weight and isospoin (whose explicit forms can be found in \cite{ET}).  The $M_{24}$ module associated with this theory via Mathieu moonshine is a graded vector space
\begin{equation}
V = \oplus_{n=1}^{\infty} V(n)
\end{equation}
with ${\rm dim} (V(n)) = A(n)$.  
Then the twining genus $Z_g$ can be written as
\begin{equation}
\label{zg}
Z_g(z;\tau) = (\chi_g - 4) {\rm ch}_{h=1/4,\ell=0}(z;\tau) - 2 {\rm ch}_{h=1/4,\ell=1/2}(z;\tau)
+ \sum_{i=1}^{\infty} A_g(n) {\rm ch}_{h=n+1/4,\ell=1/2}(z;\tau)~,
\end{equation}
with
\begin{equation}
A_g(n) = {\rm Tr}_{V(n)} g~.
\end{equation}

In practice, one can find simple closed-form expressions for $Z_g$ as discussed
in detail in e.g. \cite{MirandaTwining,GaberdielTwining,EguchiTwining}.  The first few terms in the q-expansions of the $Z_g$ for various conjugacy classes are shown in Table 3.

\begin{table}[h!]
\begin{center}
\includegraphics[width=0.9\textwidth]{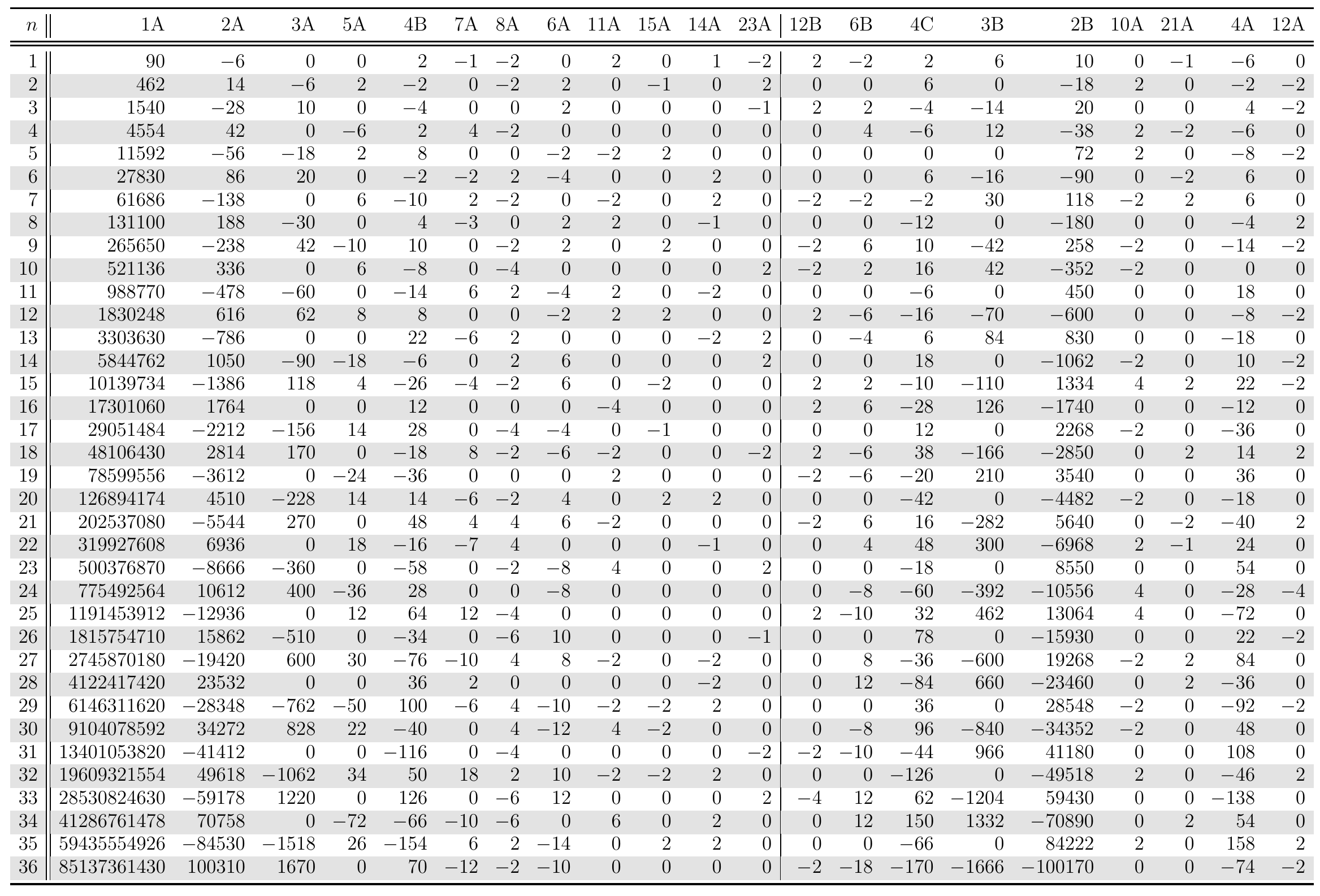}
\end{center}
\caption{Coefficients in the q-expansion of $Z_g$ for various conjugacy classes $g$.}
\end{table}

\end{document}